\documentclass[a4paper,fleqn,usenatbib]{mnras}

\usepackage[T1]{fontenc}
\usepackage{ae,aecompl}

\usepackage{graphicx}	
\usepackage{amsmath}	
\usepackage{amssymb}	
\usepackage{newtxtext,newtxmath}

\title[TIC~257060897b: an inflated,
low-density, hot-Jupiter transiting a rapidly evolving subgiant star]{TIC~257060897b: an inflated, low-density, hot-Jupiter transiting a rapidly evolving subgiant star}
\author[M. Montalto et al.]{
M. Montalto$^{1,2}$\thanks{E-mail: marco.montalto@unipd.it},
L. Malavolta$^{1,2}$,
J. Gregorio$^{3}$,
G. Mantovan$^{1,2}$,
S. Desidera$^{2}$,
\newauthor
G. Piotto$^{1,2}$,
V. Nascimbeni$^{1,2}$,
V. Granata$^{1}$,
E. E. Manthopoulou$^{1,2}$,
R. Claudi$^{2}$
\\
$^{1}$Dipartimento di Fisica e Astronomia "Galileo Galilei", Universit\'a di Padova, Vicolo dell'Osservatorio 3, Padova IT-35122, Italy\\
$^{2}$INAF Osservatorio Astronomico di Padova, vicolo dell'Osservatorio 5, 35122, Padova, Italy\\
$^{3}$Atalaia group, Crow Observatory-Portalegre, 7300 Portalegre, Portugal\\
}

\date{Accepted XXX. Received YYY; in original form ZZZ}

\pubyear{2017}

\begin{document}
\label{firstpage}
\pagerange{\pageref{firstpage}--\pageref{lastpage}}
\maketitle

\begin{abstract}
We report the discovery of a new transiting exoplanet orbiting the star TIC~257060897 and detected using {\it TESS} full frame images. We acquired HARPS-N time-series spectroscopic data, and ground-based photometric follow-up observations from which we confirm the planetary nature of the transiting body. For the host star we determined: T$\rm_{eff}$=(6128$\pm$57) K, log~g=(4.2$\pm$0.1) and [Fe/H]=(+0.20$\pm$0.04). The host is an intermediate age ($\sim$3.5~Gyr), metal-rich, subgiant star with M$_{\star}$=(1.32$\pm$0.04) M$_{\odot}$ and R$_{\star}$=(1.82$\pm$0.05) R$_{\odot}$. The transiting body is a giant planet with a mass
m$\rm_p=$(0.67$\pm$0.03) M$\rm_{j}$, a radius r$\rm_p=$(1.49$\pm$0.04) R$\rm_{j}$ yielding a density $\rho_p$=(0.25$\pm$0.02) g cm$^{-3}$ and revolving around its star every $\sim$3.66 days.
TIC~257060897b is an extreme system having one of the smallest densities known so far. 
We argued that the inflation of the planet's radius may be related to the fast increase of luminosity of its host star as it evolves outside the main sequence and that systems like TIC~257060897b could be precursors of inflated radius short period planets found around low luminosity red giant branch stars, as recently debated in the literature.
\end{abstract}




\section{Introduction}
\label{sec:introduction}

Planets orbiting subgiant stars offer the possibility to analyze a broad set of physical processes that are not at play with main-sequence stars. These phenomena range from the
atmospheric expansion and evaporation mechanisms,
the orbital period decay,
the influence of stellar mass loss on the orbital evolution of planetary systems, as well as instabilities related to the evolution of stellar binaries \citep[e.g.][]{icko1991}. Given their intermediate evolutionary state, well characterized subgiant stars can place important constraints on several physical processes that depend on the position in the Hertzsprung-Russell diagram \citep[e.g.][]{godoy2021}. 

The occurrence of planets around subgiant stars is debated in the literature. The  Lick,  Keck  and  California radial velocity planet  searches  \citep{johnson2006,johnson2007,johnson2008,johnson2010a,
johnson2010b,johnson2011a,johnson2011b}
targeted  about 500  very  bright  (V <8.5) subgiants  during  the  past  two decades and  reached  two  main  conclusions:  SG stars present (1) an higher occurrence of giant planets and (2) a lower occurrence of giant planets with short period (hot-Jupiters) with respect to main sequence planets' hosts. 
Planets found around subgiants by these surveys are usually massive (>2 M$\rm_J$) and their larger occurrence rate with respect to giant planets around main sequence stars suggests that formation of massive planets is promoted around intermediate mass subgiant stars \citep[1.5<M/M$_{\odot}$<2, e.g.][]{bowler2010}. At the same time, the lack of close-in Jupiters indicates that, during  post  main  sequence  evolution,  planets  with  orbital  separations  beyond $\sim$1  AU  are  more  likely  to  survive than planets closer to their hosts  and that orbital evolution mechanisms critically depend on the mass of the planets and of the  planets’  hosts  \citep[e. g.][]{villaver2009}. 
\citet{lillobox2016} pointed out that planetary companions to subgiant stars with semi-major axis smaller than 0.5 AU tend to be inner components of multi-planetary systems and suggested that close-in Jupiters (a<0.06 AU) are engulfed by their host stars as they evolve off the main sequence. More recently \citet{grunblatt2019} performed a systematic search of planetary transits in a sample of 2476 low luminosity red giant branch stars observed during the NASA {\it K2} mission and tentatively found a higher fraction of  inflated radius (R>1 R$\rm_{J}$) planets with short orbital periods (P<10 days) around  evolved stars than around main sequence stars. Their results suggest that close-in planets larger than Jupiter survive the subgiant phase at least until their host stars are substantially evolved (R>5-6 R$_{\odot}$).

\begin{figure}
\includegraphics[width=\columnwidth]{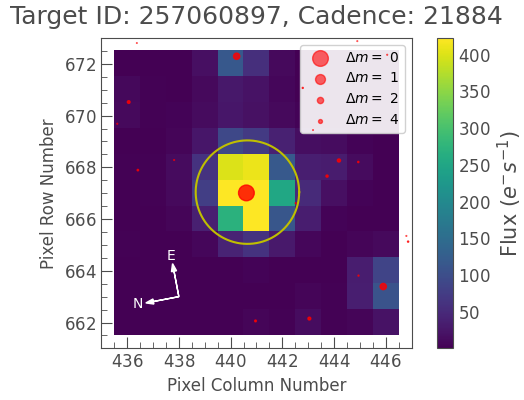}
\caption{
{\it TESS} Target Pixel File (TPF) centered on TIC~257060897 and relative to the first cadence of Sector~14. The image represents an area of 
$\sim$4.2 square arcmin around the target. Sources in {\it Gaia} EDR3 are represented by the red dots, scaled inversely proportionally to their difference of apparent $G$-band magnitude with respect to the target and corrected for proper motion to the epoch of the TPF. The yellow circle shows our adopted photometric aperture.
}
\label{fig:tesscut}
\end{figure}

\begin{figure}
\includegraphics[width=\columnwidth]{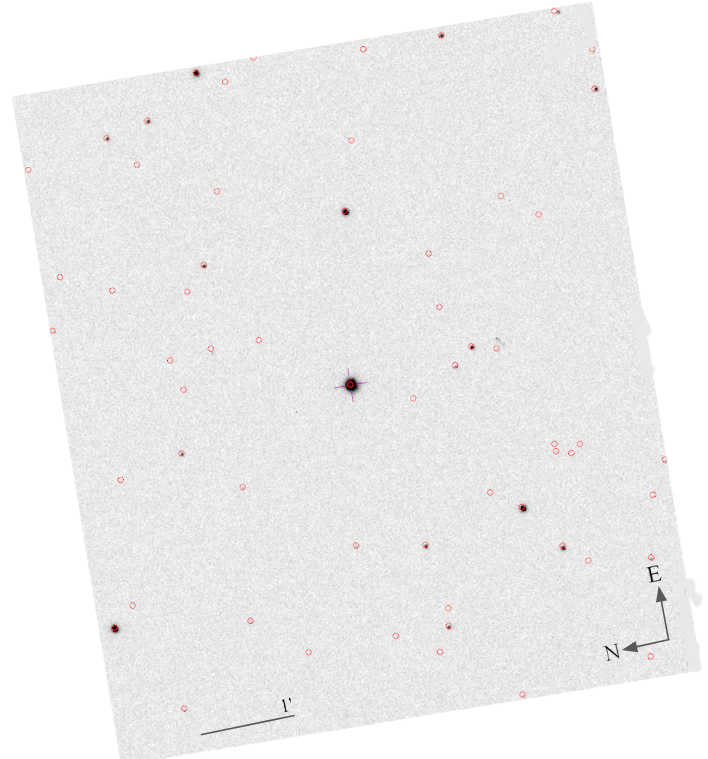}
\caption{
An area of $\sim6^{\prime}\times7^{\prime}$ centered on TIC~257060897 (indicated by the magenta cross) as imaged by the Asiago Schmidt 67/92 telescope. The image is displayed with the same orientation of the TESS image. Red open circles represent {\it Gaia} EDR3 sources.
}
\label{fig:ASIAGO_reference}
\end{figure}

The advent of the NASA {\it TESS} satellite 
\citep{ricker2015} represents an important opportunity for the study of planetary systems around evolved stars. {\it TESS} provides short-cadence (2 minutes) photometry for a sample of about 200 000 pre-selected targets across the entire sky but also delivers Full Frame images with a cadence of 30 min (during the nominal mission) and 10 minutes (during the extended mission). In \citet{montalto2020} we described a new project to exploit {\it TESS} Full Frame Images. We are monitoring a sample of about 2.6 millions FGKM dwarfs and subgiants to search for transiting planets and to globally characterize their variability properties. Subgiant stars represent nearly 50\% of this set. This is a tremendous increment of the number of evolved stars analyzed so far to search for transiting planets (by a factor of $\sim$500). We expect therefore that {\it TESS} will significantly contribute to the discovery of new planetary systems orbiting these stars. Subgiant stars are also primary targets of the next space-based planetary transits search mission PLATO \citep{rauer2014}.
In this work, we present the first discovery we achieved,
a novel short period transiting planet found around the subgiant star TIC~257060897\footnote{Recently this object has been included in the TESS Object of Interest (TOI) list as TOI~4138.01. It was alerted on June 23, 2021.}.
Table~\ref{tab:stellar_properties} summarizes some basic properties of the target star.

The procedure we followed to identify planetary transits was described in \citet{montalto2020}. We searched for planets around dwarf and subiant stars selected following the criteria described in \citet{montalto2021} using 
the box-fitting least square algorithm (BLS) of \citet{kovacs2002} and applied a random forest classifier to isolate plausible transiting planetary candidates. We also applied vetting criteria related to the centroid motion and local stellar density and inspected Gaia root-mean square radial velocity measurements whenever available to rule out obvious eclipsing binaries.

In Sect.~\ref{sec:observations}, we describe the photometric and spectroscopic observations we acquired. In Sect.~\ref{sec:data_analysis}, we describe our reduction procedure. In Sect.~\ref{sec:spectroscopic_parameters}, we explain how we determined the spectroscopic parameters of the host star, in Sect.~\ref{sec:stellar_parameters} the stellar parameters and in Sect.~\ref{sec:planetary_parameters} the planetary parameters. 
In Sect.~\ref{sec:stellar_activity}, we analyze the stellar activity.
In Sect.~\ref{sec:discussion}, we discuss our results and in Sect.~\ref{sec:conclusions} we conclude our analysis.

\begin{table}
	\centering
	\caption{Identifiers, astrometric and photometric measurements of the host star.}
	\label{tab:stellar_properties}
	\begin{tabular}{lcc} 
		\hline
		\hline
		Parameter & Value & Source \\
		\hline
	    {\it Gaia} & 1697129530714536320 & {\it Gaia} EDR3 \\
	    TYC          & TYC 4417-1588-1 & Simbad \\
	    2MASS        & J15100767+7242372 & Simbad \\
	    TIC          & 257060897 & TIC v8.1 \\
	    TOI          & 4138 & ExoFOP \\
	    $\alpha$(J2016) & 15:10:7.718 & {\it Gaia} EDR3\\
	    $\delta$(J2016) & +72:42:37.12 & {\it Gaia} EDR3\\
	    $\pi$ (mas) & 1.97$\pm$0.01 & {\it Gaia} EDR3\\
	    $\mu_{\alpha}$ (mas yr$^{-1}$) & 13.51$\pm$0.01 & {\it Gaia} EDR3\\
	    $\mu_{\delta}$ (mas yr$^{-1}$) & -7.78$\pm$0.02 & {\it Gaia} EDR3\\
	    {\it TESS}     & 11.263$\pm$0.007 & TIC v8.1\\
	    {\it G}        & 11.6617$\pm$0.0002 & {\it Gaia} EDR3\\
	    {\it G$_{BP}$} & 11.9595$\pm$0.0006 & {\it Gaia} EDR3\\
	    {\it G$_{RP}$} & 11.2016$\pm$0.0003 & {\it Gaia} EDR3\\
	    {\it B}        & 12.6$\pm$0.3 & TIC v8.1\\
	    {\it V}        & 11.81$\pm$0.02 & TIC v8.1\\
	    {\it J}        & 10.70$\pm$0.02 & TIC v8.1\\
	    {\it H}        & 10.45$\pm$0.02 & TIC v8.1\\
	    {\it K$\rm_s$} & 10.39$\pm$0.02 & TIC v8.1\\
	    {\it W1}       & 10.35$\pm$0.02 & TIC v8.1\\
	    {\it W2}       & 10.38$\pm$0.02 & TIC v8.1\\
	    {\it W3}       & 10.40$\pm$0.05 & TIC v8.1\\
	    {\it W4}       &  9.52& TIC v8.1\\
		\hline
	\end{tabular}
\end{table}

\begin{figure*}
\includegraphics[width=2\columnwidth]{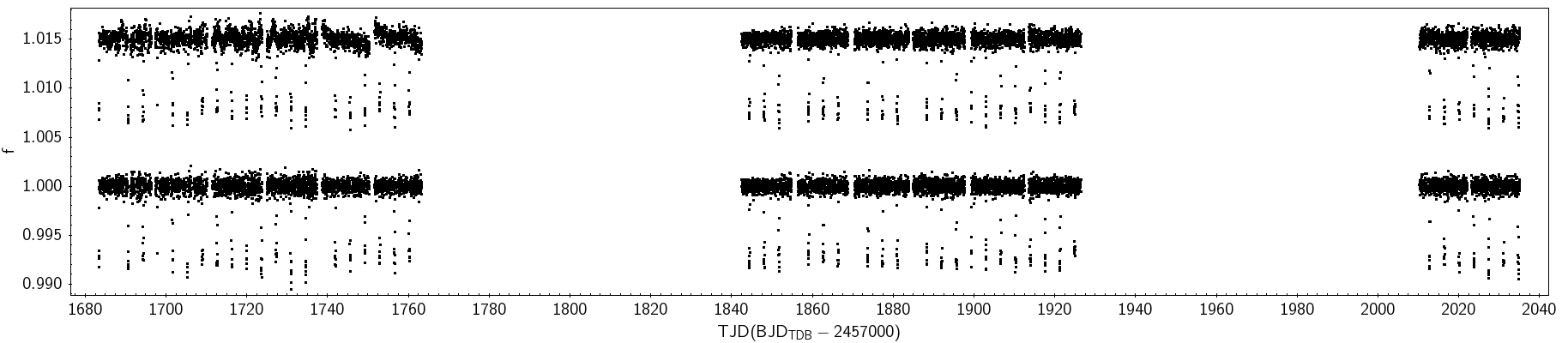}
\caption{
 {\emph Top:} {\it TESS} lightcurve corrected for systematics with eigenvector analysis. {\emph Bottom:} final lighcurve normalized by a B-spline fitted on out-of-transit data. The lightcurves are offset vertically by an arbitrary amount for clarity.
}
\label{fig:TOI257060897_full_lightcurve}
\end{figure*}

\section{Observations}
\label{sec:observations}

\subsection{Photometry}

TIC~257060897 was imaged by the {\it TESS} satellite
between July 2018 and July 2020, as reported below. We used {\it TESS} Full Frame Images to discover this object. Subsequently, we performed a ground-based follow-up using the 67/92 cm Schmidt telescope in Asiago, Italy and the 35.6 cm CROW telescope in Portalegre, Portugal.

\subsubsection{{\it TESS} photometry}

The {\it TESS} satellite imaged TIC~257060897 during the second year of operation in seven sectors: sector 14, 15, 16, 20, 21, 22 and 26. Full Frame Images were acquired with a cadence of 30 min. In total the satellite
collected 8428 images of the target. The first image was
taken on July 18, 2019 at 20:44 UT and the last image 
on July 4, 2020 at 14:43 UT. In total 47 transits have been observed. Figure~\ref{fig:tesscut} shows the Target Pixel File (TPF) for TIC~257060897 relative to the first cadence of Sector 14. 
The image represents an area of $\sim$4.2 square arcmin around the target. The red dots denote sources from {\it Gaia} EDR3 corrected for proper motion at the TPF epoch. Their dimension is scaled inversely proportionally to their difference of apparent {\it Gaia} G-band magnitude with respect to the target. We used a modified version of \texttt{tpfplotter}
\citep{aller2020} to generate this figure.

\subsubsection{Asiago 67/92 cm Schmidt telescope}

A partial transit during the egress phase was observed with the Schmidt telescope in Cima Ekar on March 2, 2021. The telescope has a correcting plate of 67 cm and a spherical mirror of 91 cm. The focal length is 215 cm. It is equipped with a KAF-16803 detector with an active area of 4096$\times$4096 pixels covering a field of view of $\sim$1 square degree with a pixel scale of 0.87 arcsec pix$^{-1}$.
The telescope is completely robotized. We used the Sloan r$^{\prime}$ filter acquiring 382 images between 18:14 UT of March 2, 2021 and 01:17 UT of March 3, 2021. We used an exposure time of 25 sec. Figure~\ref{fig:ASIAGO_reference} shows an image of the sky region with dimension $\sim7^{\prime}\times6^{\prime}$ centered on TIC~257060897 (indicated by the magenta cross) as obtained with the Asiago Schmidt 67/92 telescope.

\subsubsection{CROW Observatory, Portalegre}

The telescope is a Schmidt Cassegrain Telescope, 
Celestron C14 with aperture of 356 mm , F6 
with 2135 mm of focal length. The images were acquired with a SBIG camera model ST-10XME with CCD KAF3200ME @-20$^{\circ}$. We used the Sloan r$^{\prime}$ filter. The telescope is completely robotized and it is operated by the Atalaia group \& CROW Observatory, Portalegre, Portugal. The data were analyzed by J. Gregorio. We observed three transits with this setup. A partial transit was observed during the ingress phase between 22:44 UT on March 20, 2021 and 04:50 UT on March 21, 2021. A full transit was observed  between 20:55 UT on May 3, 2021 and 04:22 UT on May 4, 2021. 
A partial transit during egress was observed between 23:01 UT on June 5, 2021 and
02:28 on June 6, 2021.
The exposure time was fixed to 120 sec for the first and second visits and to 150 sec for the third one. A total of 90, 162 and 63 images were acquired the first, the second and the third night, respectively.

\subsection{Spectroscopy}

Spectroscopic observations were obtained with the HARPS-N \citep{cosentino2012} spectrograph\footnote{Program ID: A41TAC\_24, PI: M. ~Montalto} at the Telescopio Nazionale Galileo (TNG) at the Observatorio del Roque de los Muchachos (La Palma). We acquired 11 measurements with an exposure time of 12.6 min or 15 min obtaining a S/N$\sim$20 at 5500$\,$\AA. The measurements were acquired between May 16, 2020 and March 25, 2021, exploiting a time sharing agreement with the GAPS
({\it Global Architecture of Planetary Systems})
program \citep{covino2013,benatti2018}.

\begin{table}
	\centering
	\caption{HARPS-N radial velocities of TIC~257060897.}
	\label{tab:spectroscopic_observations}
	\begin{tabular}{ccc} 
	\hline
	\hline
        BJD$_{\textrm{TDB}}$ & RV(km s$^{-1}$) & $\rm\sigma_{RV}$ (km s$^{-1}$) \\
		\hline
        2458986.421420916  &   -11.580   &    0.005  \\
        2459026.478923034  &   -11.573   &    0.005  \\
        2459028.532127090  &   -11.719   &    0.005  \\
        2459029.430919364  &   -11.629   &    0.006  \\
        2459051.412596867  &   -11.632   &    0.005  \\
        2459072.456649624  &   -11.72    &    0.01  \\
        2459099.378091395  &   -11.589   &    0.009  \\
        2459272.722442626  &   -11.665   &    0.005  \\
        2459296.505446991  &   -11.653   &    0.005  \\
        2459297.669279696  &   -11.596   &    0.009  \\
        2459298.609597649  &   -11.69   &    0.01  \\
	\hline
        \end{tabular}
\end{table}

\begin{figure}
\includegraphics[width=0.9\columnwidth]{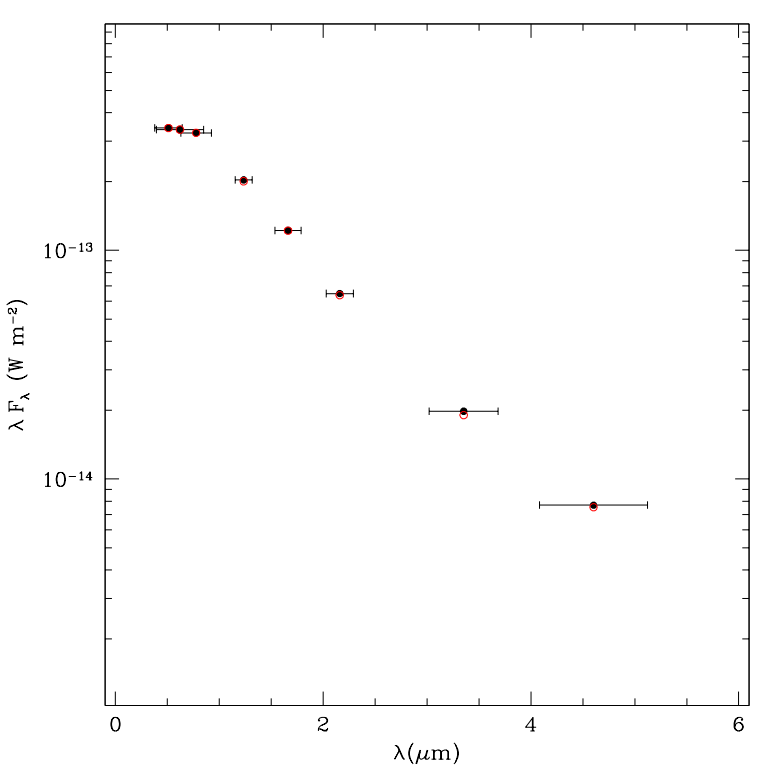}
\caption{
 Optical and near-infrared broadband photometric measurements of the target star (black circles) and best-fit model (red open circles).
}
\label{fig:SED}
\end{figure}

\begin{figure}
\includegraphics[width=\columnwidth]{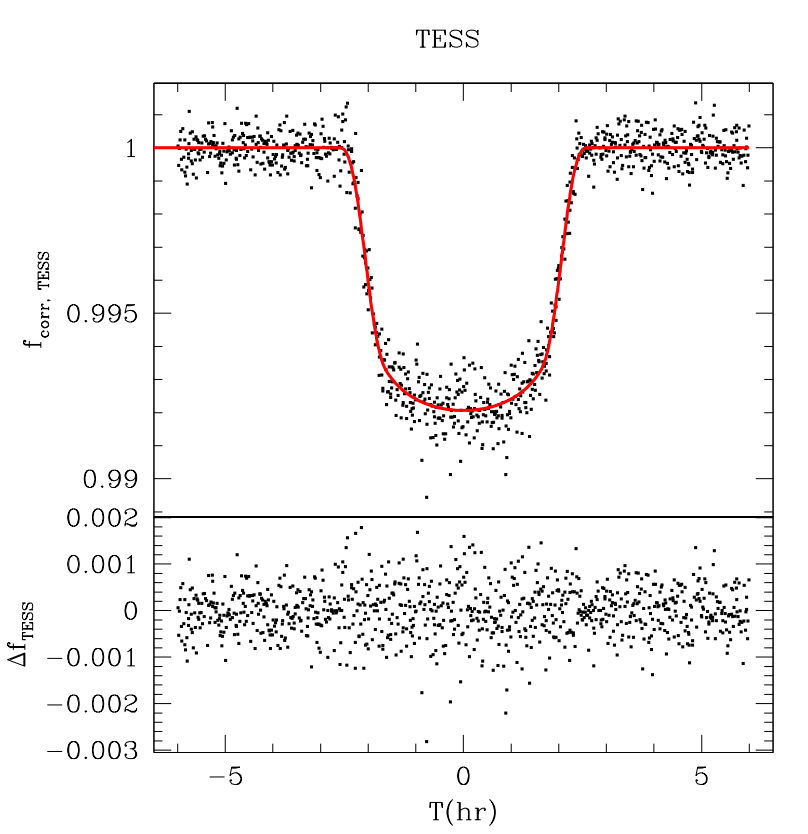}
\caption{
\emph{Top:} {\it TESS} light curve of TIC~257060897 folded with the best fit ephemerides. The best fit transit model is denoted by the red curve. \emph{Bottom:} residuals of the model fit.
}
\label{fig:TESS_LC}
\end{figure}

\begin{figure}
\includegraphics[width=\columnwidth]{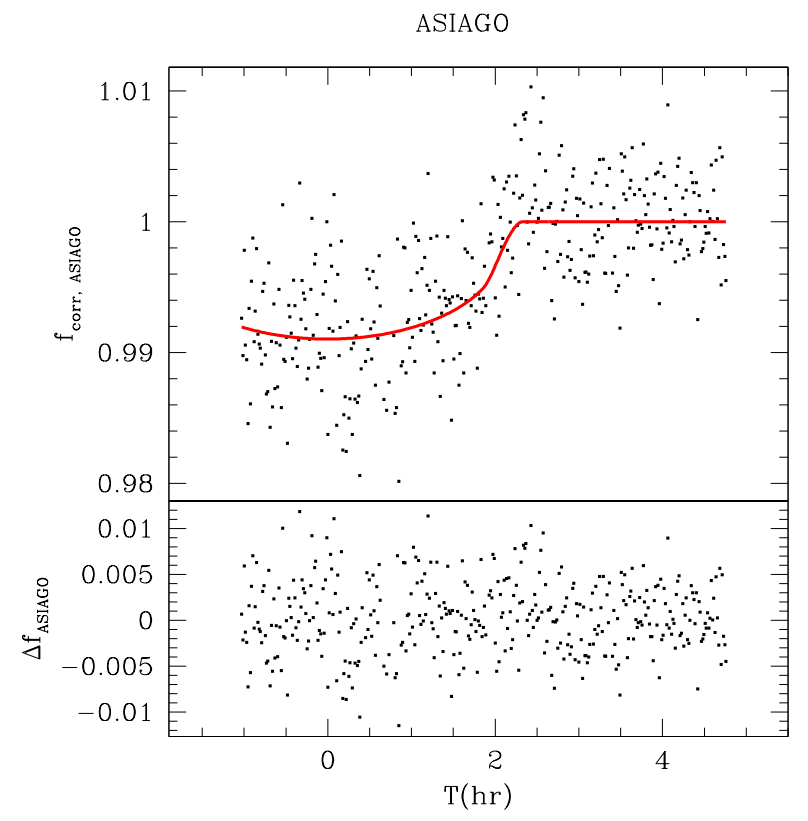}
\caption{
\emph{Top:} light curve of TIC~257060897 obtained with the Asiago 67/92 cm Schmidt telescope. The best fit transit model is 
denoted by the red curve. 
\emph{Bottom:} residuals of the model fit.
}
\label{fig:ASIAGO_LC}
\end{figure}

\begin{figure}
\includegraphics[width=\columnwidth]{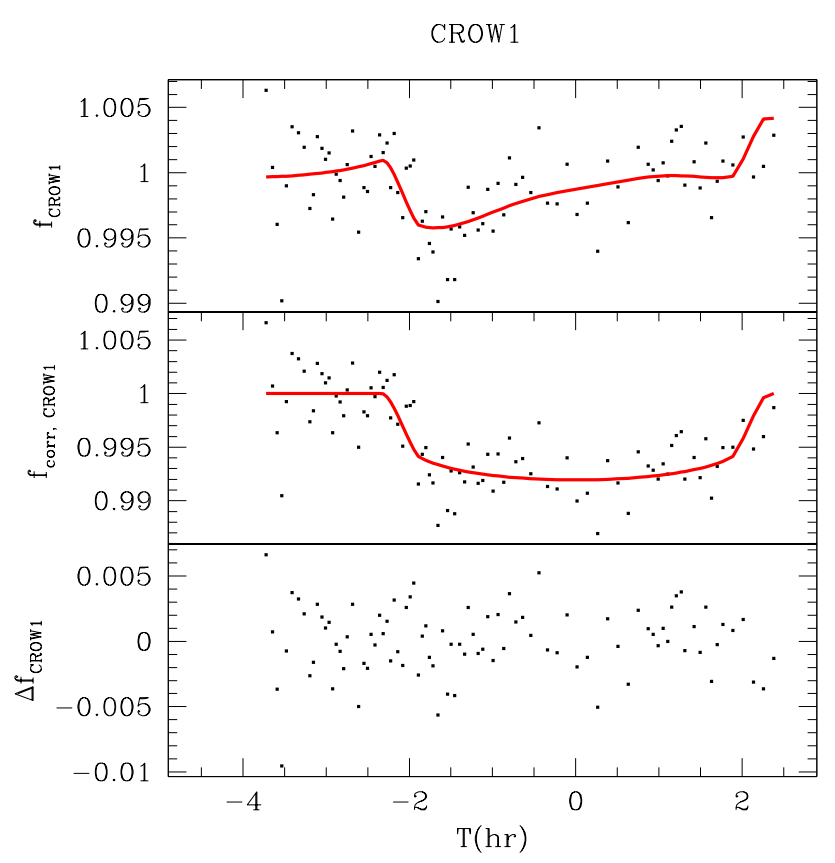}
\caption{
\emph{Top:} light curve of TIC~257060897 obtained with the CROW telescope on March 20, 2021. The best fit model (transit model + Gaussian process model) is 
denoted by the red curve. 
\emph{Middle:} light curve of TIC~257060897 
after subtracting the best fit Gaussian process model. The best fit transit model is 
denoted by the red curve. 
\emph{Bottom:} residuals of the fit.
}
\label{fig:CROW1_LC}
\end{figure}

\begin{figure}
\includegraphics[width=\columnwidth]{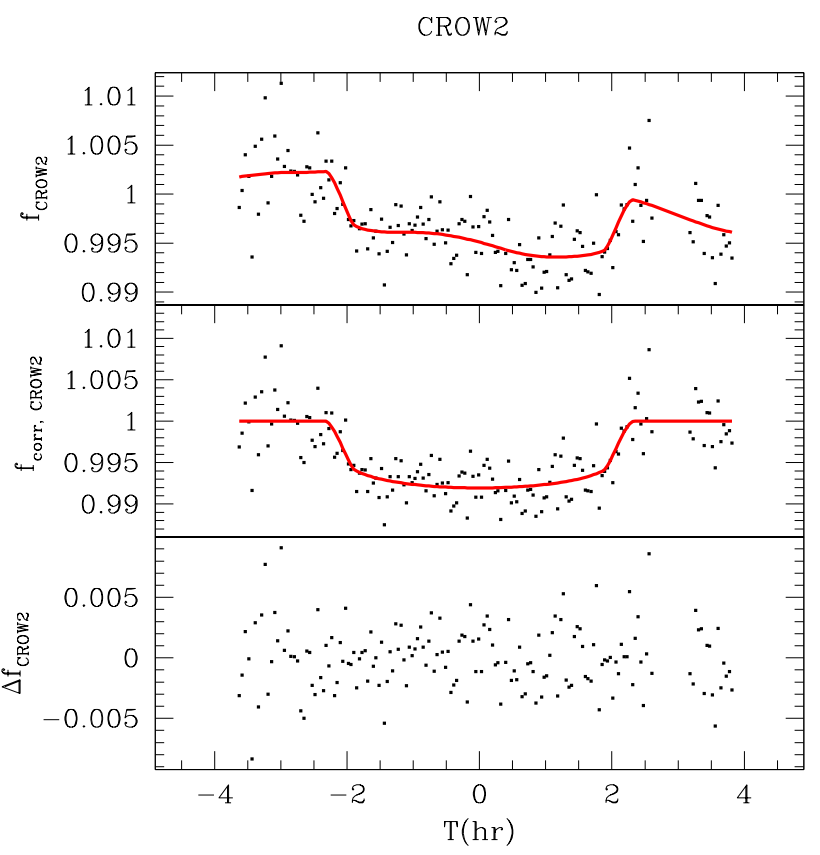}
\caption{
\emph{Top:} light curve of TIC~257060897 obtained with the CROW telescope on May 3, 2021. The best fit model (transit model + Gaussian process model) is 
denoted by the red curve.
\emph{Middle:} light curve of TIC~257060897 
after subtracting the best fit Gaussian process model. The best fit transit model is 
denoted by the red curve. 
\emph{Bottom:} residuals of the fit.
}
\label{fig:CROW2_LC}
\end{figure}

\begin{figure}
\includegraphics[width=\columnwidth]{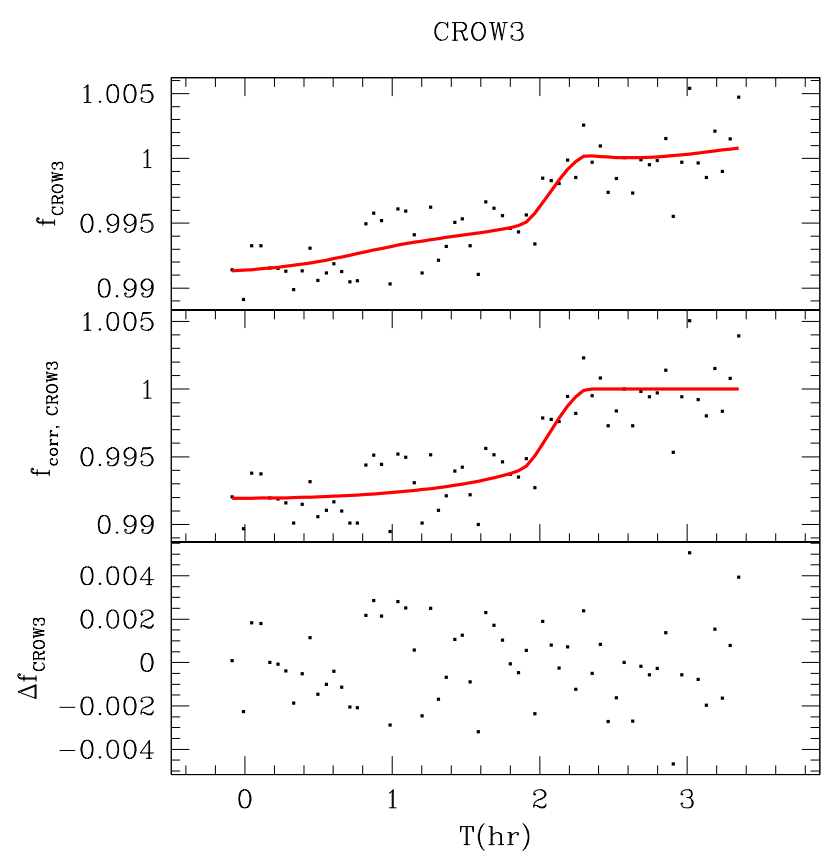}
\caption{
\emph{Top:} light curve of TIC~257060897 obtained with the CROW telescope on June 5, 2021. The best fit model (transit model + Gaussian process model) is 
denoted by the red curve.
\emph{Middle:} light curve of TIC~257060897 
after subtracting the best fit Gaussian process model. The best fit transit model is 
denoted by the red curve. 
\emph{Bottom:} residuals of the fit.
}
\label{fig:CROW3_LC}
\end{figure}

\begin{figure}
\includegraphics[width=\columnwidth]{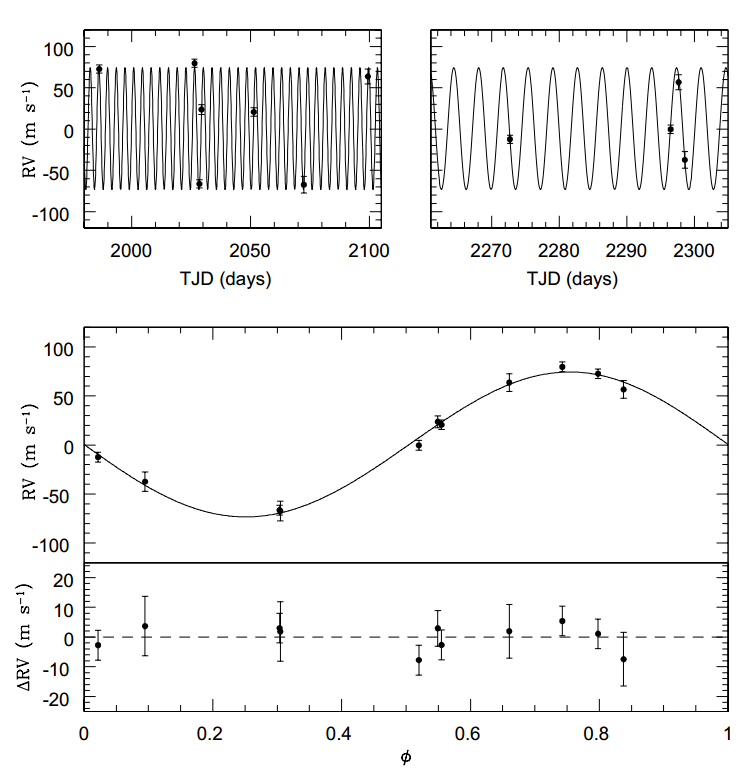}
\caption{
\emph{Top:} the two diagrams show HARPS-N radial velocity measurements as a function of time (TJD=BJD-2457000) separated in two different time intervals to better visualize them. \emph{Middle:} radial velocities folded with the best-fit ephemerides and best fit model (continuous line). \emph{Bottom:} residuals of the fit.
}
\label{fig:HARPS_RV}
\end{figure}

\section{Data analysis}
\label{sec:data_analysis}

\subsection{Photometry}

The {\it TESS} images were analyzed with the \texttt{DIAMANTE} pipeline following the procedure described in \citet{montalto2020}. In brief, the {\it TESS} FFIs were analyzed with a difference imaging approach where a stacked reference image was subtracted from each image after convolving the reference by an optimal kernel. 
The photometry was extracted using a circular aperture of radius equal to 2 pixels. We chose this aperture after testing different aperture radii between 1 pix and 4 pix.  The lightcurves from different sectors are merged together by accounting for sector by sector photometric zero points variations, then they are corrected for systematics on a sector by sector basis by using a best set of eigenlightcurves and finally normalized by a B-spline function fitted on out-of-transit data. In Fig.~\ref{fig:TOI257060897_full_lightcurve} we show the eigenvector corrected lightcurve (top) and the final B-splined lightcurve (bottom). We analyzed only the data that were not flagged by the pipeline \citep[see ][]{montalto2020} yielding 7997 measurements.

The analysis of the Asiago data was performed with custom built software. The analysis of the CROW Observatory data was done using the
AstroImageJ software\footnote{\url{https://www.astro.louisville.edu/software/astroimagej/}}. In both cases the fluxes of the target and a set of comparison stars were derived by simple aperture photometry and differential photometry was performed.

\subsection{Spectroscopy}

The spectroscopic data were reduced by the HARPS-N Data Reduction Software 
\citep[DRS v3.7, ][]{lovis2007} using a G2V template mask.  Table~\ref{tab:spectroscopic_observations} reports the radial velocities extracted by the pipeline.

\section{Spectroscopic parameters}
\label{sec:spectroscopic_parameters}

We measured effective temperature (T$\rm_{eff}$), surface gravity (log g),  iron abundance [Fe/H] and 
microturbulence velocity ($\xi$)
using the equivalent width method. We used the software \texttt{StePar} \citep{tabernero2019} which
implements a grid of MARCS model atmospheres \citep{gustafsson2008} and the MOOG \citep{sneden1973} radiative transfer code to compute stellar atmospheric parameters by means of a Downhill Simplex minimisation algorithm which minimizes a quadratic form composed of the excitation and ionisation equilibrium of Fe. Equivalent widths were measured with \texttt{ARES}~v2
\citep{sousa2015} from the  coadded spectrum obtained from the individual HARPS-N measurements used for the radial velocity measurements. The coadded spectrum had a S/N$\sim$60 at
5500 \AA. We used the FeI and FeII line list fornished by the authors for the case of the Sun.
Using this approach we obtained the parameters reported in Table~\ref{tab:spectroscopic_parameters}. 

\subsection{Empirical spectral library}
\label{sec:empirical_spectral_library}

\begin{figure}
\includegraphics[width=\columnwidth]{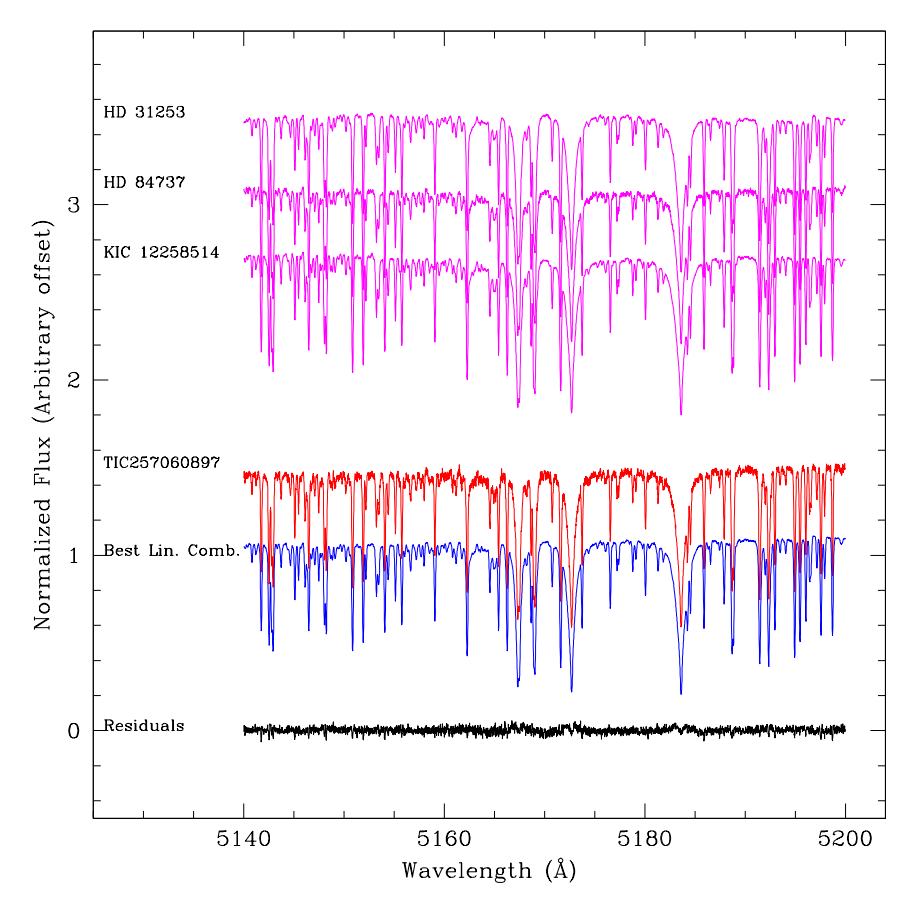}
\caption{
Comparison of the target spectrum in the Mgb triplet region with empirical
spectra in the \citet{yee2017} library. The target's spectrum is depicted in red. The three spectra of the library most highly correlated with the target's spectrum are shown in magenta, while the best-fit
linear combination of them is depicted in blue. The residuals between the best fit and the target's spectrum are represented in black. 
}
\label{fig:TIC257060897_EmpiricalLibrary.png}
\end{figure}

We also compared the spectroscopic parameters 
we derived in the previous section with the ones obtained using the spectra of the empirical library
of \citet{yee2017}. This library includes 404 stars observed with Keck/HIRES by the California Planet Search. We used the software \texttt{SpecMatch-Emp}
\citep{yee2017} to perform the comparison between our stacked spectrum of TIC~257060897 and the library spectra. In Fig.~\ref{fig:TIC257060897_EmpiricalLibrary.png} we represent with the magenta colour the three empirical library spectra\footnote{The spectra of star HD 31523, HD84737 and KIC 12258514} most highly correlated with the target spectrum
in the region of the Mgb triplet. With the red colour we indicate the target spectrum, and in blue the best-fit linear combination of the three reference spectra reported above. Finally, in black at the bottom we show the difference between the target spectrum and the linearly combined reference spectra. In this case we obtained: T$\rm_{eff}$=(5967$\pm$110) K, 
log$\,$g=(4.1$\pm$0.1) dex, [Fe/H]=(0.19$\pm$0.09) dex all consistent within 1$\sigma$ with the spectrosopic parameters previously derived and adopted.

\begin{table}
	\centering
	\caption{Spectroscopic parameters of TIC~257060897.}
	\label{tab:spectroscopic_parameters}
	\begin{tabular}{cccc} 
		\hline
		\hline
		T$\rm_{eff}$ & log$\,$g & [Fe/H] & $\rm \xi$ \\
		            (K) & (dex) & (dex) & (km s$\rm^{-1}$) \\
		\hline
		6128$\pm$57 & 4.2$\pm$0.1 & +0.20$\pm$0.04 & 1.28$\pm$0.07\\
		\hline
	\end{tabular}
\end{table}

\begin{table*}
	\centering
	\caption{System parameters relative to TIC~257060897.}
	\label{tab:system_parameters_A}
	\begin{tabular}{llccc}
		\hline
		\hline
		Parameter & Symbol & Value & Priors & Units \\
		\hline
		{\it Fitted parmeters} & & & & \\
		Transit Epoch (BJD) & $T_0$ & 1708.9983$\pm$0.0003 & $\mathcal{U}$(1708.8, 1709.1) & days \\
		Orbital Period & $P$ & 3.660028$\pm$0.000006 & $\mathcal{U}$(3.6, 3.7) & days \\
		Planet-to-star radius ratio & $p=\frac{R_p}{R_{\star}}$ & 0.0841$\pm$0.0009 & $\mathcal{U}$(0.000010; 0.5) & - \\
		Impact parameter & $b$ & 0.42$\pm$0.08 & $\mathcal{U}$(0; 2) & - \\
		Stellar reflex velocity & $K$ & 74$\pm$3 & $\mathcal{J}$(0.5; 2000) & m s$^{-1}$ \\
	    Center-of-mass velocity & $\gamma$ & -11.653$\pm$0.002 & $\mathcal{U}$(-21.720; -1.573) & km s$^{-1}$\\
	    $\sqrt{e}\cos\omega$ & $\sqrt{e}\cos\omega$ & 0.08$\pm$0.08 & $\mathcal{U}$(-1; 1) & m s$^{-1}$ \\
	    $\sqrt{e}\sin\omega$ & $\sqrt{e}\sin\omega$ & 0.0$\pm$0.2 & $\mathcal{U}$(-1; 1) & m s$^{-1}$ \\
	    Stellar density & $\rho_{\star}$ & 0.22$\pm$0.01 & $\mathcal{N}$(0.22; 0.01) & $\rho_{\odot}$\\
	    Radial velocity jitter (HARPS-N) & $\sigma_{\textrm{HARPS-N}}$ & 3$\pm$2 & $\mathcal{U}$(0.05; 1000) & m s$^{-1}$ \\
	    Jitter error (TESS) & $\sigma_{\textrm{TESS}}$ & 373$\pm$18 & $\mathcal{U}$(4, 419000) & ppm \\
	    Jitter error (ASIAGO) & $\sigma_{\textrm{ASIAGO}}$ & 1052$\pm$531 & $\mathcal{U}$(38, 389000) & ppm \\
	    Jitter error (CROW1) & $\sigma_{\textrm{CROW1}}$ & 430$\pm$356 & $\mathcal{U}$(32, 327200) & ppm \\
	    Jitter error (CROW2) & $\sigma_{\textrm{CROW2}}$ & 305$\pm$244 & $\mathcal{U}$(38, 415800) & ppm \\
	    Jitter error (CROW3) & $\sigma_{\textrm{CROW3}}$ & 1449$\pm$256 & $\mathcal{U}$(13, 132500) & ppm \\
	    GP$\rm_{\log \rho}$ (CROW) & GP$\rm_{\log \rho}$ parameter (Mat\'ern Kernel) & -2.1$\pm$0.4 & $\mathcal{U}$(-3, 3) & - \\
	    GP$\rm_{\log \sigma}$ (CROW) & GP$\rm_{\log \sigma}$ parameter (Mat\'ern Kernel) & -5.6$\pm$0.3 & $\mathcal{U}$(-6, 6) & ppm \\
	    Parameter related to linear limb darkening (TESS) & $q_{1,\textrm{TESS}}$ & 0.3$\pm$0.1 & $\mathcal{U}$(0, 1) & - \\
	    Parameter related to quadratic limb darkening (TESS) & $q_{2,\textrm{TESS}}$ & 0.3$\pm$0.2 & $\mathcal{U}$(0, 1) & - \\
	    Parameter related to linear limb darkening (ASIAGO) & $q_{1,\textrm{ASIAGO}}$ & 0.7$\pm$0.2 & $\mathcal{U}$(0, 1) & - \\
	    Parameter related to quadratic limb darkening (ASIAGO) & $q_{2,\textrm{ASIAGO}}$ & 0.5$\pm$0.3 & $\mathcal{U}$(0, 1) & - \\
	    Parameter related to linear limb darkening (CROW) & $q_{1,\textrm{CROW}}$ & 0.3$\pm$0.2 & $\mathcal{U}$(0, 1) & - \\
	    Parameter related to quadratic limb darkening (CROW) & $q_{2,\textrm{CROW}}$ & 0.4$\pm$0.3 & $\mathcal{U}$(0, 1) & - \\
	    \hline
	    {\it Derived parameters} & & & & \\
	    Orbital inclination & $i$ & 86.0$\pm$0.7 & - & $^{\circ}$\\
	    Stellar mass & $M_{\star}$ & 1.32$\pm$0.04 & - & M$_{\odot}$ \\	 
	    Stellar radius & $R_{\star}$ & 1.82$\pm$0.05 & - & R$_{\odot}$ \\	 
	    Scaled semi-major axis of the orbit  & $\frac{a}{R_{\star}}$ & 6.05$\pm$0.09 & - & - \\	 
	    Semi-major axis & a & 0.051$\pm$0.002 & - & AU \\
            Eccentricity & $e$ & 0.03$\pm$0.02 & - & - \\
	    Argument of periastron & $\omega$ & 20$\pm$72 & - & $^{\circ}$\\	   
	    Extinction in the visible & $A_{V}$ & 0.08$\pm$0.02 & - & - \\
	    Luminosity & $\log~L_{*}$ & 0.61$\pm$0.02 & - & L$\rm_{\odot}$ \\
	    Distance & $d$ & 498$\pm$13 & - & pc \\
	    Age & $\log~Age$ & 9.54$\pm$0.04 & - & \\ 
	    Planet mass &  $m_p$ &  0.67$\pm$0.03 & - & M$\rm_{jup}$\\
	    Planet radius & $r_p$  & 1.49$\pm$0.04 & - & R$\rm_{jup}$\\
	    Planet surface gravity & $log\,g_{p}$ & 2.87$\pm$0.03 & - & - \\
	    Planet density & $\rho_p$ & 0.25$\pm$0.02 & - & g cm$\rm ^{-3}$\\
	    Planet equil. temp. (A=0) & $T_{eq}$ & 1762$\pm$21 & - & K \\
	    Total duration & T$_{41}$ & 0.194$\pm$0.005 & - & days \\
	    Duration of total transit phase & T$_{32}$ & 0.158$\pm$0.006 & - & days \\
	    Linear limb darkening (TESS) & $\mu_{1,\textrm{TESS}}$ & 0.3$\pm$0.1 & - & - \\
	    Quadratic limb darkening (TESS) & $\mu_{2,\textrm{TESS}}$ & 0.2$\pm$0.3 & - & - \\
	    Linear limb darkening (ASIAGO) & $\mu_{1,\textrm{ASIAGO}}$ & 0.8$\pm$0.3 & - & - \\
	    Quadratic limb darkening (ASIAGO) & $\mu_{2,\textrm{ASIAGO}}$ & 0.0$\pm$0.4 & - & - \\
	    Linear limb darkening (CROW) & $\mu_{1,\textrm{AT1}}$ & 0.4$\pm$0.3 & - & - \\
	    Quadratic limb darkening (CROW) & $\mu_{2,\textrm{AT1}}$ & 0.1$\pm$0.3 & - & - \\	
		\hline
	\end{tabular}
\end{table*}

\begin{figure}
\includegraphics[width=\columnwidth]{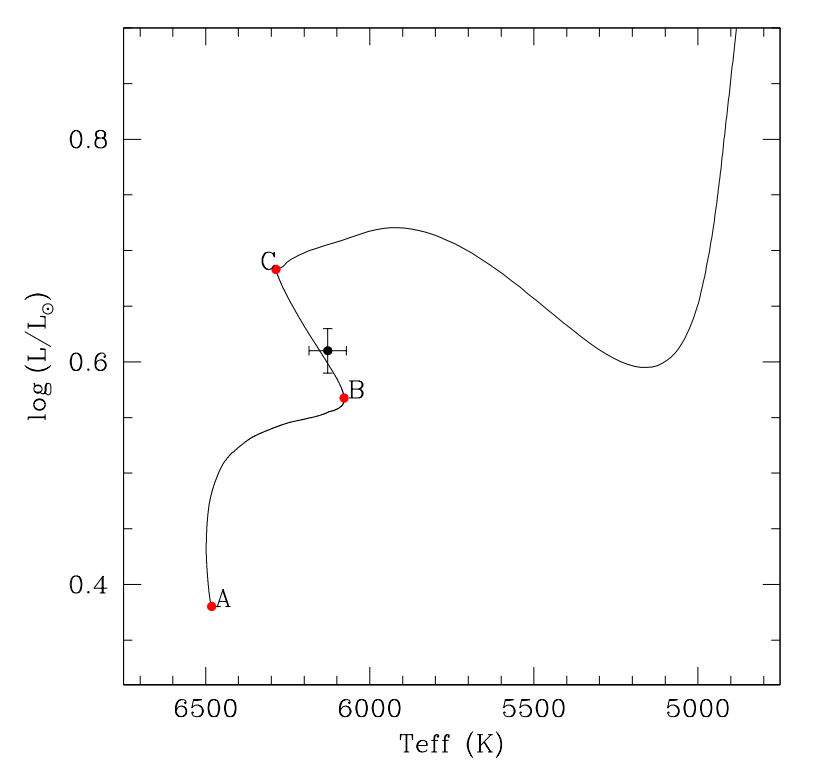}
\caption{
Stellar track of a star with mass M=1.3 M$_{\odot}$ and
metallicity [M/H]=0.2. The luminosity and the effective temperature of the target are indicated by the black dot.
The diagram shows also some critical evolutionary phases along the track (red dots). Point A: beginning of the main sequence; point B: the hydrogen burning is almost ended and a small contraction phase begins; point C: the small contraction ends, the hydrogen is exhausted in the core and the star moves towards the RGB. 
}
\label{fig:iso}
\end{figure}

\section{Stellar parameters}
\label{sec:stellar_parameters}

We derived stellar parameters using a Bayesian approach and our custom software.
In particular, we used
 the {\it Gaia} EDR3 \citep{riello2021}, 2MASS \citep{skrutskie2006, cohen2003} and ALLWISE (W1 and W2) photometry \citep{wright2010, jarrett2011}. We imposed a Gaussian prior on the effective temperature, the gravity, the metallicity using the values reported in Table~\ref{tab:spectroscopic_parameters}.
 We imposed a uniform prior on the distance: [$d$-3$\times\,ed$; $d$+3$\times\,ed$] where 
 $d$ was the distance  obtained from the simple inversion of the parallax, and $ed$ was the the semi-difference between the upper and lower estimates obtained subtracting and adding the parallax standard error to the {\it Gaia} EDR3 parallax value. 
 Such parallax value was first corrected by the zero point bias discussed in \citet{lindegren2021} using the software provided by the authors\footnote{\url{https://www.cosmos.esa.int/web/gaia/edr3-code}}. We found a value equal to -0.044953 mas
 for the bias.
 We also imposed a uniform prior on the interstellar extinction A$\rm_V$. To construct this prior we first calculated the expected value of the reddening for our target
 using the reddening map of \citet{lallement2018}. At the position of TIC257060897 we obtained {\it E(B-V)}=0.02$\pm$0.01. We assumed then a standard reddening law and obtained A$\rm_v=3.1\,E(B-V)=0.06\pm0.03$. Then we considered as plausible interval for the optical extinction the values between [0;A$\rm_v$+3$\times\sigma_{A\rm_v}$]=[0;0.15]. To calculate the expected broadband photometry we used the Padova library of stellar isochrones \citep[PARSEC,][]{bressan2012}. We first restricted the age range of the stellar isochrones to be considered within the range log Age = [9.4; 9.8] using a {\it Gaia} absolute colour-magnitude diagram. We therefore generated a set of isochrones with log Age = [9.4; 9.8] and logarithmic step size equal to 0.01 dex. We varied the metallicity of the isochrone set between [M/H] = [0.05; 0.3] in steps of size equal to 0.01 dex. For each value of the age and of the metallicity we considered nine different values of the optical extinction equal to A$\rm_V$=[0.00,0.01,0.02,0.03,0.04,0.05,0.10,0.15,0.20] and generated the corresponding models. Linear interpolation was used to derive the broadband photometry corresponding to any intermediate value of the reddening. For each value of the effective temperature, gravity, metallicity generated by the algorithm we identified the stellar model with the closest stellar parameters in our
 isochrone set. We then calculated from this model the broadband photometry applying to the model magnitudes the simulated distance modulus and extinction. We also calculated the parallax (from the simulated distance value). We then compared these simulated values of the broadband photometry and of the parallax with the observed ones. The log-likelihood function we adopted to evaluate the model performance was equal to: 
 $\ln\mathcal{L}$=-$\rm\frac{1}{2}\sum_{i=1}^{i=Nobs} (\frac{o_i-s_i}{\sigma_{o_i}})^2$. For any simulated model we registered also the value of the stellar mass, stellar radius, luminosity and age. The posterior distributions of the parameters were obtained using the nested sampling method implemented in the \texttt{MultiNest} 
package \citep{feroz2008,buchner2014,feroz2009,feroz2019}. We used 250 live points. The result of the fit is shown in Fig.~\ref{fig:SED}. All photometric data are well reproduced. The reduced chi-square of the fit ($\rm\chi_r$) is equal to $\rm\chi_r$=1.2. The best fit stellar mass and radius are M$_{\star}$=(1.32$\pm$0.04) M$_{\odot}$ and R$_{\star}$=(1.82$\pm$0.05) R$_{\odot}$, respectively. 
We also obtained a distance (d) equal to d=(498$\pm$13) pc,
an extinction equal to A$\rm_V$=(0.08$\pm$0.02) and an age
equal to log Age=(9.54$\pm$0.04).
The results of our analysis are reported in Table~\ref{tab:system_parameters_A}. In Figure~\ref{fig:iso}, we also present the stellar track of a star with mass M=1.3 M$_{\odot}$ and metallicity [M/H]=0.2. The diagram shows also some critical evolutionary phases along the track (red dots). In particular point A denotes the beginning of the main sequence (the pre-main sequence phase was neglected). At point B the hydrogen burning is almost ended. A small contraction phase begins here for intermediate and massive stars 
\citep[M$\gtrsim$1.25$\,$M$_{\odot}$, e.g.][]{kippenhahn1994}. At point C the small contraction ends, the hydrogen is exhausted in the core and the star moves toward the RGB.
The location of the target star in Figure~\ref{fig:iso} (black dot) in between points B and C suggests 
that this object has already entered a phase of instability where the hydrogen in its core is nearly completely exhausted and the core is slightly contracting before igniting the hydrogen shell. The evolution across these phases is very fast (see Sect.~\ref{sec:discussion}).

\begin{figure}
\includegraphics[width=\columnwidth]{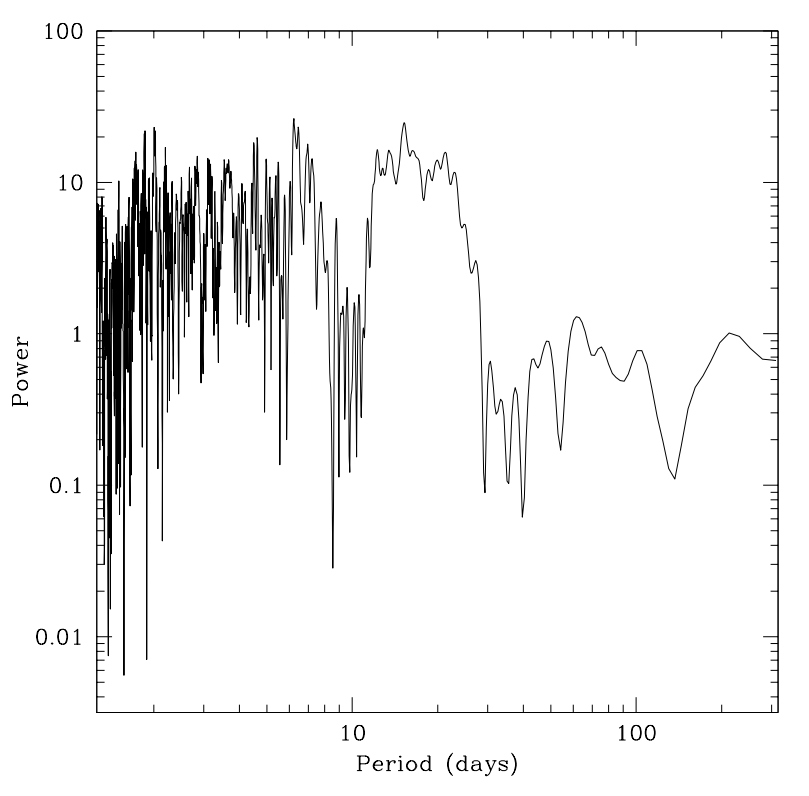}
\caption{
The Lomb-Scargle periodogram of the eigenvector corrected out-of-transit data obtained by TESS.
}
\label{fig:oot_ls}
\end{figure}

By using the empirical spectral library described in Sec.~\ref{sec:empirical_spectral_library}
we also derived the stellar parameters obtaining
R$_{\star}$=(1.7$\pm$0.2) R$_{\odot}$, M$_{\star}$=(1.20$\pm$0.08) M$_{\odot}$,  log Age=9.7, v$\,$sin$\,$i=1 km$\,$s$^{-1}$. 
The estimated v$\,$sin$\,$i and the estimated stellar radius and imply a rotation period $\sim92$ days (assuming the inclination of the stellar rotation axis is identical to the inclination of the planetary orbit).
The Lomb-Scargle periodogram of the eigenvector corrected out-of-transit data obtained by TESS is rather flat beyond $\sim$30 days as shown in Fig.~\ref{fig:oot_ls}, whereas some structure is visible for smaller periods although it is not associated with a clear modulation.

An alternative method to determine the stellar parameters is described in \citet{montalto2021} and employed for the construction of the all-sky PLATO input catalogue (asPIC1.1). 
In this case we obtained
T$\rm_{eff}$=(5946$\pm$254) K,
R$_{\star}$=(2.0$\pm$0.2) R$_{\odot}$ and M$_{\star}$=(1.3$\pm$0.1) M$_{\odot}$.
These estimates are all compatible with the parameters obtained by the bayesian approach described above, which was finally adopted in our analysis.

\section{Planetary parameters}
\label{sec:planetary_parameters}

Planetary parameters were obtained performing a simultaneous fit
of both spectroscopic and photometric data with the software
\texttt{PyORBIT} \citep{malavolta2016,malavolta2018}.
For the transiting planet we assumed a Keplerian orbit with free eccentricity, following the parametrization of \citet{eastman2013}.
Transit models were computed with the \texttt{batman} package \citep{kreidberg2015}, following the prescriptions of \citet{benatti2019}.
For TESS long-cadence data, we took into account the distortion due to the extended integration time \citet{kipping2010} by averaging the light curve model over ten evenly-spaced points computed within each 1800s exposure. For the other datasets, this correction was not deemed necessary.

We used independent limb-darkening quadratic coefficients for each instrument. We sampled limb-darkening  coefficients with the method described in \citet{kipping2013} and used uninformative priors for all parameters except the stellar density, for which we used a Gaussian prior following the results of Sect.~\ref{sec:stellar_parameters}. Each dataset came with its own jitter parameter to absorb underestimated white-noise errors and unaccounted sources of red noise, with each CROW transit treated as an independent dataset.

We explored the possibility of modelling instrumental systematic effects in the light curves with a Gaussian processes (GP) using a Matérn-like kernel, as implemented in the code \texttt{celerite} \citep{ambikasaran2014,foreman2017}.
We performed model selection among different combinations of datasets with/without GP by computing the Bayesian evidence through Dynamic Nested Sampling  \citep{higson2019}, after implementing \texttt{dynesty} \citep{speagle2020} into \texttt{PyORBIT}. The favourite model foresees the use of a GP for CROW data only, with hyper parameters shared among the three transits. This model was moderately favoured over the use of independent hyper parameters for each CROW transit ($\Delta \ln\mathcal{Z} = 4.9 \pm 0.8$) or an additional GP for the Asiago light curve ($\Delta \ln\mathcal{Z} = 4.2 \pm 0.8$), while it was strongly favoured over any other combination  ($\Delta \ln\mathcal{Z} > 10 $). Regarding radial velocities, we preferred to leave any possible activity-related signal to be absorbed by the jitter parameter, rather than using a GP, due to the relatively small number of observations.

The posterior distributions of the parameters were obtained using \texttt{emcee} \citep{foremanmackay2013}, employing the most favourite model according to the Bayesian evidence. The model encompassed 23 parameters, for which we used 92 MCMC walkers. We run the MCMC for 100000 steps, conservatively discarding the first 25000 steps as burn-in and applying a thinning factor of 100. The confidence interval were estimated by taking the 15.86th and 84.13th percentiles of the posterior, reported together with the median values in Table~\ref{tab:system_parameters_A}. We obtained that the transiting body is a Jupiter-like planet with a mass m$\rm_p=$(0.67$\pm$0.03) M$\rm_{j}$ and a radius r$\rm_p=$(1.49$\pm$0.04) R$\rm_{j}$ yielding a density $\rho_p$=(0.25$\pm$0.02) g cm$^{-3}$. The resulting eccentricity is equal to $e=$(0.03$\pm$0.02) consistent with a circular orbit according to the criterion of \citep{lucy1971}. The best fit model is represented in the Figures \ref{fig:TESS_LC}-\ref{fig:HARPS_RV}.

\section{Stellar activity}
\label{sec:stellar_activity}

\subsection{Chromospheric activity}
\label{sec:chromospheric_activity}

In Fig.~\ref{fig:bis_logR} (top) we present the chromospheric activity index log R$^{\prime}\rm_{HK}$ as a function of the radial velocity measurements. To calculate this index we first calculated the $S\rm_{CaII}$ index with the software \texttt{ACTIN}
\footnote{https://github.com/gomesdasilva/ACTIN}
\citep{dasilva2018} and then followed the procedure reported in
\citet{dasilva2021} to calibrate the $S\rm_{CaII}$ index to the Mount Wilson scale and to calculate the photospheric and bolometric
corrected chromospheric emission ratio R$^{\prime}\rm_{HK}$. Errors on the individual measurements are obtained from error propagation of the formulas reported in \citet{dasilva2021}\footnote{We also considered a minimum error on Mount Wilson in dex of 0.0005 dex,
as suggested in \citet{dasilva2021}.}. The Pearson correlation coefficient between the log R$^{\prime}\rm_{HK}$ index and the radial velocities is equal to 0.23 (p-value=0.5416) indicating a negligible correlation between these two quantities. The average log R$^{\prime}\rm_{HK}$ is 
<log R$^{\prime}\rm_{HK}$>=(-5.06$\pm$0.05).  According to the classification proposed by \citet{henry1996}, during our observations
TIC~257060897 was inactive, consistently with the moderately old age derived from stellar models in Sect.~\ref{sec:stellar_parameters}.

\subsection{Bisector span}
\label{sec:bisector_span}

\begin{figure}
\includegraphics[width=\columnwidth]{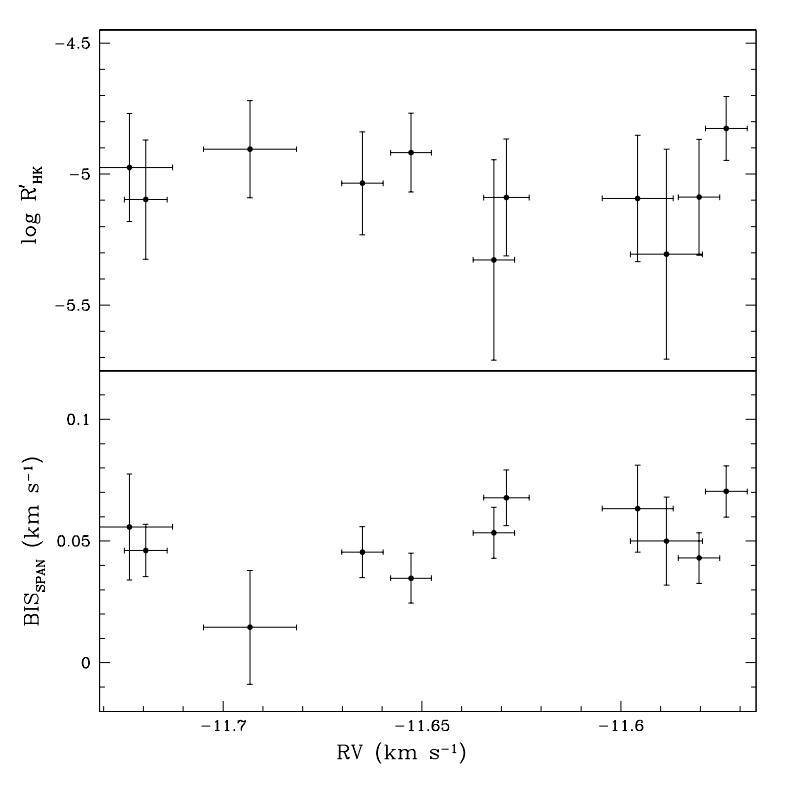}
\caption{
\emph{Top: } the chromospheric activity index of TIC~257060897 vs the radial velocities measurements. \emph{Bottom: } the bisector span vs the radial velocities.
}
\label{fig:bis_logR}
\end{figure}

In Fig.~\ref{fig:bis_logR} (bottom) we report the bisector span \citep{queloz2001} vs the radial velocity measurements. Such quantity was calculated by the HARPS-N pipeline and it is a measure of the line asymmetry which may as well suggest the presence of issues related to activity and/or multiplicity.  The error on the bisector has been assumed equal to twice the error on the radial velocities. Also in this case we measured a negligible correlation between the radial velocities and the bisector span (r$\rm_{pearson}$=0.44, p-value=0.1808).

\begin{figure}
\includegraphics[width=\columnwidth]{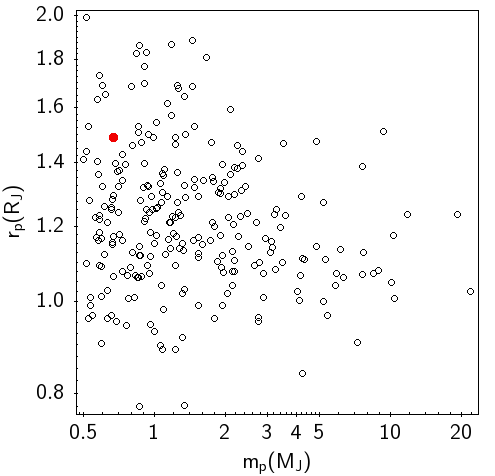}
\caption{
Planetary radius vs planetary mass for the 250 known transitng planets with mass>0.5 M$\rm_{J}$ and with masses and radii measured with a precision better than 10$\%$ (open circles). The position of TIC~257060897b in this
diagram is indicated by the red dot.
}
\label{fig:radius_mass}
\end{figure}

\begin{figure}
\includegraphics[width=\columnwidth]{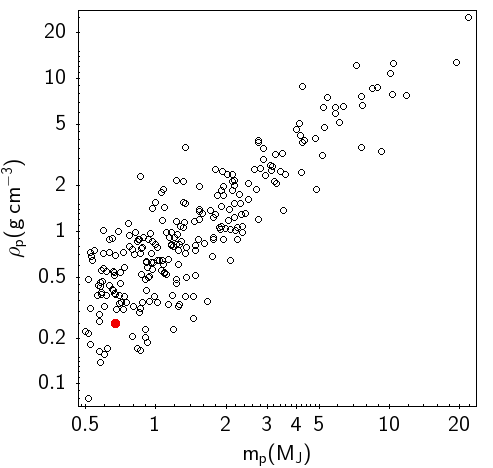}
\caption{
Planetary density vs planetary mass  for the same sample of planets presented in Fig.~\ref{fig:radius_mass}. The position of TIC~257060897b in this
diagram is indicated by the red dot.
}
\label{fig:density_mass}
\end{figure}

\section{Discussion}
\label{sec:discussion}

The lack of correlation between RVs vs log R$^{\prime}\rm_{HK}$ and RVs vs BIS$\rm_{span}$
supports the Keplerian origin of the RV variations
and the planetary nature of the transiting object.
TIC~257060897b is an inflated hot-Jupiter planet with a very low density and surface gravity. In 
Fig.~\ref{fig:radius_mass}, we present the 
radius against the mass of all 250 known transiting exoplanets (with mass>0.5 M$\rm_{J}$) having a precision better than 10$\%$ in both radius and mass\footnote{This list was retrieved from the NASA exoplanet archive: \url{http://exoplanetarchive.ipac.caltech.edu} on September 21, 2021. Whenever a planet had multiple entries in the table we averaged the quantities of interest among the different entries.} and in Fig.\ref{fig:density_mass} the density vs the mass for the same planets' sample. From these figures, it is clear that TIC~257060897b is one of the largest radius and smallest density transiting exoplanets known so far.
This result contributes to make TIC~257060897b an extreme planetary system in the context of known hot-Jupiter planets.
The inflated radius of TIC~257060897b 
may be put in relation with the evolutionary state of its host star. Since the beginning of the main sequence this star increased its luminosity by 70$\%$ in about 3.5 Gyr. Over the next 130 Myr it will increase further its luminosity by 30$\%$. Therefore TIC~257060897 is in a phase of extremely rapid luminosity increment (see Fig.~\ref{fig:iso}) and the The atmosphere of the close-in exoplanet may have reacted puffing up in response of the increment of energy input from the host. 

\begin{figure}
\includegraphics[width=\columnwidth]{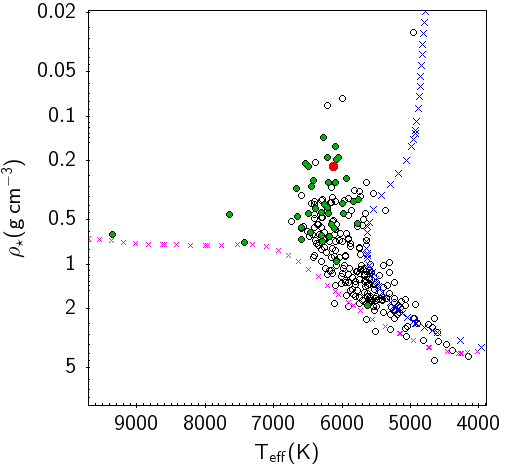}
\caption{
Stellar density vs effective temperature diagram for the same sample of planets presented in Fig.~\ref{fig:radius_mass}.
The position of TIC~257060897b in this diagram is indicated by the red dot, while inflated Jupiter planets with radius  $>$1.45 R$\rm_{J}$ are represented by the green dots. Magenta and blue crosses show a 100 Myr and a 13 Gyr solar metallicity isochrone from the Padova database.
}
\label{fig:InflatedJupiters}
\end{figure}

\begin{figure}
\includegraphics[width=\columnwidth]{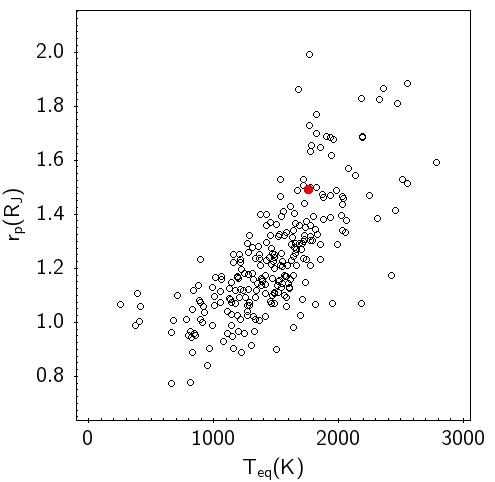}
\caption{
Planetary radius vs planetary equilibrium temperature diagram for the same sample of planets presented in Fig.~\ref{fig:radius_mass}. The position of TIC~257060897b in this diagram is indicated by the red dot.
}
\label{fig:RadiusTeq}
\end{figure}

The argument of re-inflation of close-in giant planets has been extensively discussed
in \citet{hartman2016} who demonstrated that inflated radius
Hot-Jupiters are preferentially found around more evolved host stars and that
this is not due to any kind of observational bias. TIC~257060897b is a new
object that belongs to the class of inflated planets around moderately evolved
stars and therefore appears to naturally support the idea of re-inflation.
In Figure~\ref{fig:InflatedJupiters}, we show that inflated radii planets (R$>$1.45 R$\rm_J$, green dots) are found preferentially
around the most evolved stars also in the sample we analyzed.
As discussed in \citet{hartman2016} this is likely a consequence of
the well established correlation between the equilibrium
temperature and planetary radius, once accounting for the fact that
planets around evolved stars are generally more highly irradiated than planets
around main sequence stars (at the same orbital distance). TIC~257060897b appears to follow such a correlation (Fig.~\ref{fig:RadiusTeq}).
 The important theoretical implication of re-inflation is that the
incident energy should be deposited deep in the interiors of the planets in order to permit the rapid expansion
of the planetary atmosphere \citep{lopez2016}  and present theories of radius inflation are still not able to convincingly explain
how this could happen irrespectively of the mechanism that it is considered
\citep[e.g.][]{thorngren2018}. Probably re-inflation is fornishing an important clue on where to look to further
improve theory. This is especially true if the sample of inflated planets around moderately evolved stars will
continue to grow in the future, which also demonstrates the importance of targeting subgiant and giant stars
to discover new planets around them \citep{lopez2016}.

Considering the mass of the host, the fate of this planetary system is likely to be engulfed in the stellar envelope \citep{villaver2009}, but the exact time of engulfment depends on several factors among which the mass, the internal structure of the planet and its initial orbit around the parent star seem to play a crucial role \citep[e.g.][]{villaver2014}. 
 The inflated radius of TIC~257060897b and the possibility that systems like this one may be capable to survive at least up to the base of the red-giant branch \citep{villaver2014} suggest that they may be also precursors of super-Jupiter planets at short orbital period around low luminosity red giant branch stars, as recently discussed in the literature \citep{grunblatt2019}.

The low density and high equilibrium temperature of 
TIC~257060897b make this object an attractive target in the contex of exoplanet atmospheric studies.
The planet's equilibrium temperature ($T_{eq}$)
was calculated assuming zero albedo and full day-night heat redistribution according to

\begin{equation}
T_{eq} = T_{\ast}\sqrt{\frac{R_{\ast}}{a}}\Big(\frac{1}{4}\Big)^{1/4}
\end{equation}

\noindent
where $a$ is the orbital semi-major axis given in the same units as $R_{\ast}$ (the stellar radius) and  $T_{\ast}$ is the host star effective temperature.
Assuming a molecular hydrogen (H$_2$)-dominated, cloud-free atmosphere, we can estimate the scale height of the planetary atmosphere, 
H=$\rm\frac{k_b\,T_{eq}}{\rm\mu\,g}$ where $k_b$ is the Boltzmann constant, $T\rm_{eq}$ is the planet equilibrium temperature, $\mu$ is the mean molecular mass and $g$ is the gravity. We obtain 
H = (985 $\pm$ 70) km. The amplitude of spectral features in trasmission is $\rm\sim4pH/R\rm_{s}=(182\pm14) $~ppm \citep{kreidberg2018}, where p is the radius ratio and R$\rm_{s}$ is the radius of the star.
We also calculated the Transmission Spectroscopy Metric 
\citep[TSM; ][]{kempton2018} a parameter which quantifies the expected signal-to-noise in transmission spectroscopy for a given planet and permits to determine its suitability for future atmospheric characterization in particular using the James Webb Space Telescope (JWST). The analytic expression for this parameter is:

\begin{equation}
\mbox{TSM} = S \times \frac{R_p^3T_{eq}}{M_pR_{\ast}^2} \times 10^{-m\rm_J/5}
\end{equation}

\noindent
where S is a normalization constant to match the more detailed work of
\citet{louie2018}, $R_p$ is the radius of the planet in units of Earth radii, $M_p$
is the mass of the planet in units of Earth masses, $R_{\ast}$ is the radius of the host
star in units of solar radii, $m\rm_J$ is the apparent magnitude of the host star in the J band and T$_{eq}$ is expressed in Kelvin. Following \citet{martin2021} we decided to choose the scale factor S = 1.15. For TIC 257060897b we found a value of 
TSM = $98\pm11$. Jupiter and sub-Jupiter planets with TSM values greater than 90 are considered suitable for transmission spectroscopy observations with JWST and TIC257060897b belongs to the 70$\rm^{th}$ percentile of the cumulative distribution of TSM values of the planets
we considered in Figg.~\ref{fig:radius_mass}-~\ref{fig:RadiusTeq}. Moreover, with an ecliptic latitude of 74.415$^{\circ}$, TIC 257060897b is near the northern JWST continuous viewing zone and will be observable for at least 197 continuous days per year \citep{gardner2006}.

Finally, we note that the host star of TIC~257060897b is also metal-rich. The Jupiter planet frequency-metallicity correlation \citep[e.g.][]{santos2004, fischer2005} predicts that Jupiter-like planets should be preferentially found around metal-rich stars. This fact is considered as a proof in favour of the core-accretion planet formation model \citep[e.g.][]{pollack1996} and TIC~257060897b therefore may have been formed by means of this mechanism in the outer stellar disk to then migrate inward at the location where we now observe it.

\section{Conclusions}
\label{sec:conclusions}

We report the discovery of TIC~257060897b, an inflated, low-density hot-Jupiter orbiting a rapidly evolving subgiant star and detected using {\it TESS} Full Frame Images. We performed photometric and spectroscopic ground-based follow-up observations which permit to conclude that the host star is an intermediate age ($\sim$3.5 Gyr), metal-rich, subgiant star with T$\rm_{eff}$=(6128$\pm$57) K, log~g=(4.2$\pm$0.1) and [Fe/H]=(0.20$\pm$0.04) implying M$_{\star}$=(1.32$\pm$0.04) M$_{\odot}$ and R$_{\star}$=(1.82$\pm$0.05) R$_{\odot}$. 
The transiting body is a giant planet with mass m$\rm_p=$(0.67$\pm$0.03) M$\rm_{J}$, radius r$\rm_p=$(1.49$\pm$0.04) R$\rm_{J}$ yielding a density $\rho_p$=(0.25$\pm$0.02) g cm$^{-3}$
and orbiting its star every $\sim$3.66 days.
TIC~257060897b is one of the hot-Jupiters with the smallest density known so far. It is also an excellent target for
atmospheric characterization with the James Webb Space Telescope. We suggest that the inflated radius of this object may be related 
to the fast increase of luminosity of its host star as itevolves outside the main sequence
and that systems like TIC~257060897b could be precursors of  inflated radius short period planets found around low luminosity
red giant branch stars, as recently debated in the literature.

\section*{Acknowledgements}
Based on observations made with the Italian {\it Telescopio Nazionale Galileo} (TNG) operated by
the {\it Fundaci\'on Galileo Galilei} (FGG) of the {\it Istituto Nazionale di Astrofisica} (INAF) at
the {\it  Observatorio del Roque de los Muchachos} (La Palma, Canary Islands, Spain) and
on observations collected at the Schmidt telescope (Asiago, Italy) of the INAF - Osservatorio
Astronomico di Padova. We thank the GAPS collaboration for the time sharing agreement and for handling the scheduling
and execution of the observations. MM is grateful to the TNG staff for the prompt support during the preparation
and execution of the observations. MM, GP, VN, VG, RC acknowledge support from PLATO ASI-INAF
agreements n.2015-019-R0-2015 and n. 2015-019-R.1-2018. This work made use of \texttt{tpfplotter} by
J. Lillo-Box (publicly available in \url{http://www.github.com/jlillo/tpfplotter}), which also
made use of the python packages \texttt{astropy}, \texttt{lightkurve}, \texttt{matplotlib} and \texttt{numpy}. 

\section{Data Availability}
The light curves and spectroscopic data presented in this  article  are available online as electronic tables.




\bibliographystyle{mnras}
\typeout{}
\bibliography{paper} 

\bsp	
\label{lastpage}
\end{document}